# Soft Decision Cooperative Spectrum Sensing Based Upon Noise Uncertainty Estimation

Hossam M. Farag
Dept. of Electrical Engineering, Aswan University, Egypt.
hossam.farag@aswu.edu.eg

Ehab Mahmoud Mohamed
Current: Dept. of Information & Communication Technology, Osaka University, Japan.
Permanent: Dept. of Electrical Engineering, Aswan University, Egypt.
ehab@wireless.comm.eng.osaka-u.ac.jp

*Abstract*—Spectrum Sensing (SS) constitutes the most critical task in Cognitive Radio (CR) systems for Primary User (PU) detection. Cooperative Spectrum Sensing (CSS) is introduced to enhance the detection reliability of the PU in fading environments. In this paper, we propose a soft decision based CSS algorithm using energy detection by taking into account the noise uncertainty effect. In the proposed algorithm, two threshold levels are utilized based on predicting the current PU activity, which can be successfully expected using a simple successive averaging process with time. The two threshold levels are evaluated based on estimating the noise uncertainty factor. In addition, they are toggled in a dynamic manner to compensate the noise uncertainty effect and to increase the probability of detection and decrease the probability of false alarm. Theoretical analysis is performed on the proposed algorithm to evaluate its enhanced false alarm and detection probabilities over the conventional soft decision CSS using different combining schemes. In addition, simulation results show the high efficiency of the proposed scheme compared to the conventional soft decision CSS, with high computational complexity enhancements.

## I. INTRODUCTION

Cognitive Radio (CR) technique is a feasible solution to solve the conflicts between spectrum scarcity and spectrum underutilization [1]. It allows the Cognitive Radio users (CRs) to share the spectrum with the Primary User (PU) by opportunistic accessing. Spectrum sensing (SS) is the most critical issue in the CR technology, since it needs to detect the presence and the absence of the PU accurately and swiftly; high efficient SS scheme will result in highly operated CR node. Common methods of SS are energy detection, cyclostationary detection and filter matching detection [2]. Energy detection is the most preferable approach for SS due to its simplicity and applicability as it does not need any prior knowledge about the PU signal. Thus, it can be applied to any communication system [2].

Cooperative Spectrum Sensing (CSS) is a promising strategy to enhance the sensing performance of the Cognitive Radio Network (CRN). In this scheme, multiple CRs are cooperated to detect the presence and the absence of the PU in order to combat some sensing problems such as fading, shadowing and the CR hidden node problem [3]. In energy detection based CSS, each CR individually measures the energy received from the PU, and the final decision is calculated based on the Fusion Center (FC) rule. In hard decision CSS [4], each CR individually decides the current PU activity based on its own measurements. Then, it sends its final decision to the FC, which in turn applies the AND, OR or Majority rule on the collected CRs decisions to arrive at a final decision about the current PU activity. On the other hand, in soft decision CSS, the CRs work as energy sensors for the FC. They frequently send their own PU energy measurements to the FC. Then, the FC applies one of the energy combining schemes on the collected energy measurements, such as Square-Law Combining (SLC), Maximal Ratio Combining (MRC) and Square-Law Selection (SLS) to arrive at a final decision about the current PU activity [5]. Soft decision CSS has higher performances than hard decision CSS; this is the reason why we only focus on soft decision CSS in this paper.

A lot of theoretical research work has been done to study the effect of noise uncertainty in the energy detection based SS [6] - [8]. At the best of our knowledge, most of the current research work only studied the effect of noise uncertainty using single PU-CR configuration. No proposal has been suggested to investigate the noise uncertainty effect using energy detection based CSS. We believe that our proposal is one of the early research efforts to suggest an efficient soft decision CSS, which overcomes the noise uncertainty effect.

In this paper, we propose a practical energy detection based soft decision CSS algorithm to improve the sensing performance of the CRN by overcoming the noise uncertainty effect. In the proposed scheme, to combat the noise uncertainty effect, the FC stores the combined values of the CRs energy measurements in addition to the average values of their estimated noise variances for a predefined period. By using these combined energy values, the FC can predict, to some extent, the presence or the absence of the PU in the current sensing event, thanks to the fact that the time required by a PU to change his status (ON/OFF) is negligible to the time the PU remains at a certain status (ON/OFF). In this paper, we focus on the main combining schemes commonly used in soft decision CSS, which are SLC, MRC, and SLS. Based on predicting the current PU activity, a new decision threshold is set to maximize the detection probability, if the PU is predicted to be present or to minimize the false alarm probability if the PU is predicted to be absent. The two dynamic thresholds are evaluated using an estimated value of the CRs noise uncertainty factor, which can be calculated using the average values of the CRs estimated noise variances over the predefined period. Theoretical and simulation analysis show that the proposed scheme outperforms the conventional soft decision CSS, in which noise uncertainty effect is not considered, using the same number of CRs. In addition, a lower number of CRs are required by the proposed scheme to obtain the same performance like the conventional soft decision CSS.

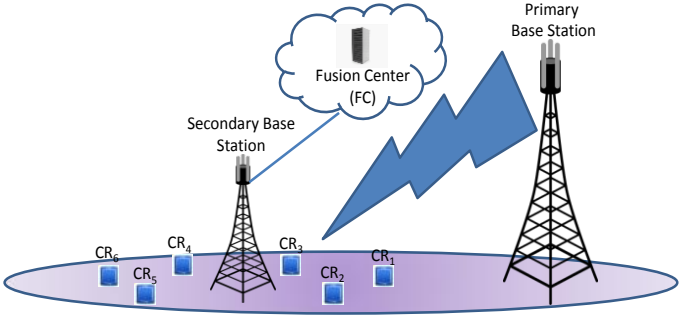

Fig. 1. The Cognitive radio network (CRN) architrcure.

The rest of this paper is organized as follows: Section II reviews the conventional soft decision CSS. Section III describes the proposed algorithm for soft decision CSS. Section IV provides the theoretical analysis of the proposed scheme. Performance evaluations are given in Sect. V. Complexity analysis is given in Sect. VI, followed by the conclusion in Sect. VII.

## II. THE CONVENTIONAL SOFT DECISION CSS

Figure 1 shows a typical CRN architecture, which we will consider throughout this paper. In this CRN, one secondary base station (BS) including the FC and $K$ CR users are located in the area of one primary base station (PU). For the sake of simplicity, noiseless reporting channels are assumed between the CRs and the secondary BS. The received signal at the $j$-th CR user from the PU, $j=1, 2,…, K,$ can be defined as:

$$y(n)_j = \begin{cases} w(n)_j & H_0 \\ h_j s(n) + w(n)_j & H_1 \end{cases}, \quad (1)$$

$n=1, 2,..., N,$ where $N$ is the total number of received samples, $y(n)_j$ is the received signal at the $j$-th CR, and $w(n)_j$ is the Additive White Gaussian Noise (AWGN) at the $j$-th CR. $s(n)$ is the PU signal, and $h_j$ is the complex channel response between the PU and the $j$-th CR; $h_j$ will be assumed as a Rayleigh flat fading channel. Hypothesis $H_0$ states that PU is absent, and hypothesis $H_1$ states that PU is present. Using energy detector, the received energy $E_j$ at the $j$-th CR is given by:

$$E_j = \sum_{n=1}^{N} y(n)_j^2. \quad (2)$$

CSS takes the advantage of spatial diversity to improve the spectrum sensing performance against fading and shadowing problems. In soft decision CSS, the detection of the PU is achieved by combining all individual sensing information of CRs at the FC using various combining schemes like SLC, MRC, and SLS [5].

The SLC is one of the simplest soft decision schemes as no channel estimation is needed [9]. In this scheme, the measured energy $E_j$ of each CR is sent to the FC, where they will be added together. Then, the result is compared with a predefined threshold to determine the presence of the PU. The combined energy of the SLC is given as [5]:

$$E_{SLC} = \sum_{j=1}^{K} E_j. \quad (3)$$

For SLC, the false alarm probability $Q_{fa,SLC}$ and the detection probability $\overline{Q}_{d,SLC}$ can be calculated as follows [5]:

$$Q_{fa,SLC} = P_r(E_{SLC} \geq \lambda \setminus H_0) = \frac{\Gamma\left(Ku, \frac{\lambda}{2\sigma^2}\right)}{\Gamma(Ku)}, \quad (4)$$

$$\overline{Q}_{d,SLC} = P_r(E_{SLC} \geq \lambda \setminus H_1) = \int_0^\infty Q_{d,SLC}^{AWGN} f(\gamma_{SLC}) \, d\gamma_{SLC}, \quad (5)$$

where $\Gamma(.,.)$ and $\Gamma(.)$ are the incomplete and complete gamma functions, respectively. $u$ is the time - bandwidth product, and $\lambda$ is the decision threshold. $\sigma^2$ is the noise variance, and $\gamma_{SLC} = \sum_{j=1}^{K}\gamma_j$, where $\gamma_j$ is the Signal-to-Noise-Ratio (SNR) of the $j$-th CR. $Q_{d,SLC}^{AWGN}$ is the detection probability of SLC under AWGN channel, and $f(\gamma_{SLC})$ is the probability density function (PDF) of $\gamma_{SLC}$ under Rayleigh fading [5].

$$Q_{d,SLC}^{AWGN} = Q_{Ku}\left(\sqrt{\frac{2\gamma_{SLC}}{\sigma^2}}, \sqrt{\frac{\lambda}{\sigma^2}}\right), \quad (6)$$

$$f(\gamma_{SLC}) = \frac{1}{\Gamma(K)}\left(\frac{1}{\overline{\gamma}}\right)^k \gamma_{SLC}^{(k-1)} e^{-\frac{\gamma_{SLC}}{\overline{\gamma}}}, \quad \gamma_{SLC} \geq 0, \quad (7)$$

where $Q_{Ku}(.,.)$ is the generalized Marcum $Q$-function, and $\overline{\gamma}$ is the average SNR.

In MRC, the received energy value $E_j$ by the FC is multiplied by a weight value $w_j$, which is proportional to $\gamma_j$, and then added together. The combined energy of the MRC is given as [5]:

$$E_{MRC} = \sum_{j=1}^{K} w_j E_j, \quad w_j = \frac{\gamma_j}{\sum_{j=1}^{K}\gamma_j}. \quad (8)$$

The false alarm probability $Q_{fa,MRC}$ and the detection probability $\overline{Q}_{d,MRC}$ for MRC can be expressed as [5]:

$$Q_{fa,MRC} = P_r(E_{MRC} \geq \lambda \setminus H_0) = \frac{\Gamma\left(u, \frac{\lambda}{2\sigma^2}\right)}{\Gamma(u)}, \quad (9)$$

$$\overline{Q}_{d,MRC} = P_r(E_{MRC} \geq \lambda \setminus H_1) = \int_0^\infty Q_{d,MRC}^{AWGN} f(\gamma_{MRC}) \, d\gamma_{MRC}, \quad (10)$$

where $\gamma_{MRC} = \sum_{j=1}^{K}\gamma_j$, $Q_{d,MRC}^{AWGN}$ is the detection probability of MRC under AWGN channel, and $f(\gamma_{MRC})$ is the PDF of $\gamma_{MRC}$ under Rayleigh fading [5]:

$$Q_{d,MRC}^{AWGN} = Q_u\left(\sqrt{\frac{2\gamma_{MRC}}{\sigma^2}}, \sqrt{\frac{\lambda}{\sigma^2}}\right), \quad (11)$$

$$f(\gamma_{MRC}) = \frac{1}{\overline{\gamma}^K (K-1)!} \gamma_{MRC}^{(k-1)} e^{-\frac{\gamma_{MRC}}{\overline{\gamma}}}, \quad \gamma_{MRC} \geq 0. \quad (12)$$

In SLS, the FC only selects the highest energy CR, i.e., $E_{SLS} = \max_{1 \leq j \leq K}(E_j)$. The SLS false alarm probability $Q_{fa,SLS}$ and detection probability $\overline{Q}_{d,SLS}$ are given as follows [5]:

$$Q_{fa,SLS} = P_r(E_{SLS} \geq \lambda \setminus H_0) = 1 - \left[1 - \frac{\Gamma\left(u, \frac{\lambda}{2\sigma^2}\right)}{\Gamma(u)}\right]^K, \quad (13)$$

$$\overline{Q}_{d,SLS} = P_r(E_{SLS} \geq \lambda \setminus H_1) = \int_0^\infty Q_{d,SLS}^{AWGN} f(\gamma_j) \, d\gamma_j, \quad (14)$$

where $Q_{d,SLS}^{AWGN}$ is the detection probability of SLS under AWGN channel, and $f(\gamma_j)$ is the PDF of $\gamma_j$ under Rayleigh fading [5].

$$Q_{d,SLS}^{AWGN} = 1 - \prod_{j=1}^{K}\left[1 - Q_u\left(\sqrt{\frac{2\gamma_j}{\sigma^2}}, \sqrt{\frac{\lambda}{\sigma^2}}\right)\right], \quad (15)$$

$$f(\gamma_j) = \frac{1}{\overline{\gamma}} e^{-\frac{\gamma_j}{\overline{\gamma}}}, \quad \gamma_j \geq 0. \quad (16)$$

In the conventional soft decision CSS, using different FC combining schemes, the effect of noise uncertainty in the CRs energy measurements is never taken into account or compensated by the FC. Consequently, the performance of the conventional soft decision CSS will be limited by the famous phenomenon named the SNR wall [7].

## III. THE PROPOSED SOFT DECISION CSS

In this section, an efficient soft decision CSS technique is proposed. The proposed technique takes into account the noise uncertainty effect when arriving at a final decision about the current PU activity. This can be simply done by storing the CRs successive energy measurements in addition to their estimated noise variances for a specific period ($L$). Current PU activity (presence/absence) can be predicted using these stored energy measurements. By predicting the current PU activity, a new dynamic threshold is toggled between two levels to compensate the noise uncertainty effect and to maximize the probability of detection/minimize the probability of false alarm if the PU is currently predicted to be present/to be absent, respectively. The two threshold levels are evaluated using an estimated value of the noise uncertainty factor.

Therefore, in the proposed scheme, a PU activity profile is created by storing the past $L$-1 combined energy values $E_{comb}(i) \in \{E_{SLC}(i), E_{MRC}(i), E_{SLS}(i)\}$, $1 \leq i \leq L-1$, of $K$ CRs, at the FC. In addition, a noise variance history is created by storing the $L$-1 average values of the $K$ CRs noise variances, $\sigma_{mean}^2(i)$, which can be calculated as:

$$\sigma_{mean}^2(i) = \frac{1}{K}\sum_{j=1}^{K}\sigma_j^2(i), \quad 1 \leq i \leq L-1. \quad (17)$$

Based on these stored combined energy values, $E_{comb}(i)$, the FC can predict, to some extent, the current PU status. The FC can do that by averaging the $L$ combined energy values $E_{comb}(i)$ including the combined energy of the current observation event $E_{comb}(L)$, as follows:

$$E_{avg} = \frac{1}{L}\sum_{i=1}^{L}E_{comb}(i). \quad (18)$$

At the same time, using $\sigma_{mean}^2(i)$ including the current observation event $\sigma_{mean}^2(L)$, the noise uncertainty factor $\rho$ of the whole energy measurement process can be estimated by the FC as follows [7], [8]:

$$\rho = \frac{\max_{1\leq i \leq L}\left(\sigma_{mean}^2(i)\right)}{\frac{1}{L}\sum_{i=1}^{L}\sigma_{mean}^2(i)}. \quad (19)$$

After calculating $E_{avg}$ and $\rho$, the FC can predict, to some extent, the current PU status by comparing $E_{avg}$ with a predefined threshold $\lambda$ which is calculated using the Constant False Alarm Rate (CFAR) approach as follows:

$$\lambda = \begin{cases} \sigma^2\left(erfc^{-1}(2Q_{fa,SLC})(2\sqrt{2Ku}) + 2Ku\right), & \text{For } SLC \\ \sigma^2\left(erfc^{-1}(2Q_{fa,MRC})(2\sqrt{2u}) + 2u\right), & \text{For } MRC \\ \sigma^2\left(erfc^{-1}\left(2(1-(1-Q_{fa,SLS})^{1/k})\right)(2\sqrt{2u}) + 2u\right), & \text{For } SLS \end{cases}, u \gg 1. \quad (20)$$

The $\lambda$ values in (20) are calculated using Gaussian approximations to (4), (9) and (13) [10].

Thanks to the fact that the time required by a PU to change his status (ON/OFF) is negligible to the time the PU remains at a certain status (ON/OFF), if $E_{avg} \geq \lambda$, the FC can expect, to some extent, the presence of the PU in the current PU activity $L$. Accordingly, a new decision threshold $\lambda_{new}$ for the final decision is set to equal $\lambda/\rho$, which results in maximizing the overall probability of detection by reducing the noise uncertainty effect. On the other hand, if $E_{avg} < \lambda$, the FC can expect, to some extent, the absence of the PU in the current observation period $L$, so $\lambda_{new}$ is set to equal $\rho\lambda$, which minimizes the overall probability of false alarm by combating the noise uncertainty effect. Hence, $\lambda_{new}$ is calculated as:

$$\lambda_{new} = \begin{cases} \lambda/\rho & \text{if } E_{avg} \geq \lambda \\ \rho\lambda & \text{if } E_{avg} < \lambda \end{cases}. \quad (21)$$

The final decision at the current sensing event $L$ becomes:
$$\begin{aligned} E_{comb}(L) \geq \lambda_{new} & \quad (PU \text{ is present}) \\ E_{comb}(L) < \lambda_{new} & \quad (PU \text{ is absent}) \end{aligned}. \quad (22)$$

With this simple mechanism of dynamic threshold selection, the proposed algorithm is expected to have a significantly improved performance over the conventional soft decision CSS using the same number of CRs, as we will prove using theoretical and simulation analysis. In addition, a lower number of CRs are expected to be used by the proposed scheme to obtain the same performance as the conventional one. Thus, the complexity of the sensing process is highly decreased using the proposed scheme.

## IV. THEORETICAL ANALYSIS OF THE PROPOSED SCHEME

In this section, we prove the effectiveness of the proposed scheme over the conventional soft decision CSS via mathematical derivations. In the theoretical analysis, we mainly focus of the false alarm and the detection probabilities.

### A. Using SLC

According to the Central Limit Theorem (CLT), for sufficiently large $N$ ($N \gg 1$), the received energy $E_j$ can be approximated as Gaussian [11]:

$$E_j \sim \begin{cases} \mathcal{N}(N\sigma^2, 2N\sigma^4) & H_0 \\ \mathcal{N}(N\sigma^2(\gamma_j+1), 2N\sigma^4(\gamma_j+1)^2) & H_1 \end{cases}. \quad (23)$$

According to (23), $E_{comb}(i)$ can be also assumed as Gaussian. Hence, (4) and (6) can be re-written as [10]:

$$Q_{fa,SLC}(\lambda) \cong Q\left(\frac{\lambda - NK\sigma^2}{\sigma^2\sqrt{2NK}}\right), \quad (24)$$

$$Q_{d,SLC}^{AWGN}(\lambda) \cong Q\left(\frac{\lambda - NK\sigma^2(\gamma_{SLC}+1)}{\sigma^2(\gamma_{SLC}+1)\sqrt{2NK}}\right), \quad (25)$$

where $Q(.)$ is the standard Gaussian complementary cumulative distribution function, and $\lambda$ is given in (20). Since $E_{avg}$ in (18) is the average of independent and identically distributed (i.i.d.) Gaussian random variables, it is also considered to be normally distributed:

$$E_{avg} \sim \mathcal{N}(\mu_{avg}, \sigma_{avg}^2). \quad (26)$$

For the SLC scheme, $\mu_{avg}$ and $\sigma_{avg}^2$ can expressed as:

$$\mu_{avg} = \frac{M}{L} NK\sigma^2(\gamma_{SLC}+1) + \frac{L-M}{L} NK\sigma^2, \quad (27)$$

$$\sigma_{avg}^2 = \frac{M}{L^2} 2NK\sigma^4(\gamma_{SLC}+1)^2 + \frac{L-M}{L^2} 2NK\sigma^4, \quad (28)$$

where $M \in \{0, L\}$ is the number of sensing events where PU is actually present. Based on the proposed algorithm, for AWGN channel, the false alarm probability and the detection probability using SLC becomes:

$$\begin{aligned} Q_{fa,SLC}^{Prop} &= \{P_r(E_{avg} \geq \lambda, E_{comb}(L) \geq \lambda/\rho)\}_{H_0} + \{P_r(E_{avg} < \lambda, E_{comb}(L) \geq \rho\lambda)\}_{H_0} \\ &= \{P_r(E_{comb}(L) \geq \lambda/\rho \setminus E_{avg} \geq \lambda) \cdot P_r(E_{avg} \geq \lambda)\}_{H_0} \\ &+ \{P_r(E_{comb}(L) \geq \rho\lambda \setminus E_{avg} < \lambda) \cdot P_r(E_{avg} < \lambda)\}_{H_0} \end{aligned} \quad (29)$$

$$\begin{aligned} Q_{d,SLC}^{Prob,AWGN} &= \{P_r(E_{avg} \geq \lambda, E_{comb}(L) \geq \lambda/\rho)\}_{H_1} + \{P_r(E_{avg} < \lambda, E_{comb}(L) \geq \rho\lambda)\}_{H_1} \\ &= \{P_r(E_{comb}(L) \geq \lambda/\rho \setminus E_{avg} \geq \lambda) \cdot P_r(E_{avg} \geq \lambda)\}_{H_1} \\ &+ \{P_r(E_{comb}(L) \geq \rho\lambda \setminus E_{avg} < \lambda) \cdot P_r(E_{avg} < \lambda)\}_{H_1} \end{aligned} \quad (30)$$

$E_{avg}$ is calculated over a representative number ($L$) of stored combined energy values $E_{comb}(i)$ to accurately predict the current PU activity. Since the average of a relatively large set of values is not significantly affected, in general, by the value of a single element, it is reasonable to assume that $E_{avg}$ is approximately independent of $E_{comb}(i)$ for sufficiently large $L$. Consequently, we get:

$$P_r(E_{comb}(L) \geq \lambda/\rho \setminus E_{avg} \geq \lambda) \cong P_r(E_{comb}(L) \geq \lambda/\rho), \quad (31)$$

$$P_r(E_{comb}(L) \geq \rho\lambda \setminus E_{avg} < \lambda) \cong P_r(E_{comb}(L) \geq \rho\lambda), \quad (32)$$

using (31) and (32), (29) and (30) become:

$$\begin{aligned} Q_{fa,SLC}^{Prop} &\cong \{P_r(E_{avg} \geq \lambda)\}_{H_0} \cdot \{P_r(E_{comb}(L) \geq \lambda/\rho)\}_{H_0} \\ &+ \{P_r(E_{avg} < \lambda)\}_{H_0} \cdot \{P_r(E_{comb}(L) \geq \rho\lambda)\}_{H_0} \end{aligned} \quad (33)$$

$$\begin{aligned} Q_{d,SLC}^{Prob,AWGN} &\cong \{P_r(E_{avg} \geq \lambda)\}_{H_1} \cdot \{P_r(E_{comb}(L) \geq \lambda/\rho)\}_{H_1} \\ &+ \{P_r(E_{avg} < \lambda)\}_{H_1} \cdot \{P_r(E_{comb}(L) \geq \rho\lambda)\}_{H_1} \end{aligned} \quad (34)$$

Based on (24) and (26), the proposed false alarm probability $Q_{fa,SLC}^{Prop}$ can be calculated as:

$$\begin{aligned} Q_{fa,SLC}^{Prop} &\cong Q\left(\frac{\lambda - \mu_{avg}}{\sigma_{avg}}\right) \cdot Q_{faSLC}(\lambda/\rho) + \left[1 - Q\left(\frac{\lambda - \mu_{avg}}{\sigma_{avg}}\right)\right] \cdot Q_{faSLC}(\rho\lambda) \\ &\cong Q\left(\frac{\lambda - \mu_{avg}}{\sigma_{avg}}\right) \cdot \left[Q_{faSLC}(\lambda/\rho) - Q_{faSLC}(\rho\lambda)\right] + Q_{faSLC}(\rho\lambda) \end{aligned} \quad (35)$$

Based on (34), the detection probability in Rayleigh fading channel conditions $\overline{Q_{d,SLC}^{Prop}}$ can be calculated as:

$$\overline{Q_{d,SLC}^{Prop}} = \int_0^\infty Q_{d,SLC}^{Prob,AWGN} f(\gamma_{SLC}) \, d\gamma_{SLC}. \quad (36)$$

Using (25) and (26), $Q_{d,SLC}^{Prob,AWGN}$ in (34) can be calculated as:

$$\begin{aligned} Q_{d,SLC}^{Prob,AWGN} &\cong Q\left(\frac{\lambda - \mu_{avg}}{\sigma_{avg}}\right) \cdot Q_{d,SLC}^{AWGN}(\lambda/\rho) + \left[1 - Q\left(\frac{\lambda - \mu_{avg}}{\sigma_{avg}}\right)\right] \cdot Q_{d,SLC}^{AWGN}(\rho\lambda) \\ &\cong Q\left(\frac{\lambda - \mu_{avg}}{\sigma_{avg}}\right) \cdot \left[Q_{d,SLC}^{AWGN}(\lambda/\rho) - Q_{d,SLC}^{AWGN}(\rho\lambda)\right] + Q_{d,SLC}^{AWGN}(\rho\lambda) \end{aligned} \quad (37)$$

Given that the PU is actually absent ($H_0$) during the period $L$, a reliable PU activity predictor will get $Q\left(\frac{\lambda - \mu_{avg}}{\sigma_{avg}}\right) \cong 0$. With such a value, from (35), we get $Q_{fa,SLC}^{Prop} \cong Q_{faSLC}(\rho\lambda)$, so $Q_{fa,SLC}^{Prop}$ will be lower than the conventional $Q_{faSLC}(\lambda)$ given by (24). By analogy, given $H_1$, we will get $Q\left(\frac{\lambda - \mu_{avg}}{\sigma_{avg}}\right) \cong 1$, so from (37), we get $Q_{d,SLC}^{Prob,AWGN} \cong Q_{d,SLC}^{AWGN}(\lambda/\rho)$, which is greater than $Q_{d,SLC}^{AWGN}(\lambda)$ given by (25). Therefore, $\overline{Q_{d,SLC}^{Prop}}$ in (36) will be higher than the conventional $\overline{Q}_{d,SLC}$ in (5).

The PU activity predictor will inaccurately predict the current PU status at the time when the PU changes his status from ON to OFF and vice versa. This will increase the probability of false alarm and decrease the probability of detection of the proposed scheme compared to the conventional method. But, these cases will not affect the overall improved performance of the proposed scheme because the time required by a typical PU to change his status (ON/OFF) is negligible to the time the PU remains at a certain status (ON/OFF). Therefore, using the same number of CRs, the proposed soft decision CSS will have a better performance than the conventional soft decision CSS using SLC.

*B. Using MRC*

Similarly, using MRC, based on the approximated distribution in (23), (9) and (11) can be expressed as [10]:

$$Q_{fa,MRC}(\lambda) \cong Q\left(\frac{\lambda - N\sigma^2}{\sigma^2\sqrt{2N}}\right), \quad (38)$$

$$Q_{d,MRC}^{AWGN}(\lambda) \cong Q\left(\frac{\lambda - N\sigma^2(\gamma_{MRC}+1)}{\sigma^2(\gamma_{MRC}+1)\sqrt{2N}}\right). \quad (39)$$

For (26), $\mu_{avg}$ and $\sigma_{avg}^2$ for MRC can be calculated as:

$$\mu_{avg} = \frac{M}{L} N\sigma^2(\gamma_{MRC}+1) + \frac{L-M}{L} N\sigma^2, \quad (40)$$

$$\sigma_{avg}^2 = \frac{M}{L^2} 2N\sigma^4(\gamma_{MRC}+1)^2 + \frac{L-M}{L^2} 2N\sigma^4. \quad (41)$$

Similar to the SLC case, (29)-(34), and based on (26) (38) (39), the overall false alarm probability $Q_{fa,MRC}^{Prop}$ and the overall detection probability $\overline{Q_{d,MRC}^{Prop}}$ for the proposed MRC scheme can be expressed as follows:

$$\begin{aligned} Q_{fa,MRC}^{Prop} &\cong Q\left(\frac{\lambda - \mu_{avg}}{\sigma_{avg}}\right) \cdot Q_{fa,MRC}(\lambda/\rho) + \left[1 - Q\left(\frac{\lambda - \mu_{avg}}{\sigma_{avg}}\right)\right] \cdot Q_{fa,MRC}(\rho\lambda) \\ &\cong Q\left(\frac{\lambda - \mu_{avg}}{\sigma_{avg}}\right) \cdot \left[Q_{fa,MRC}(\lambda/\rho) - Q_{fa,MRC}(\rho\lambda)\right] + Q_{fa,MRC}(\rho\lambda) \end{aligned} \quad (42)$$

$$\overline{Q_{d,MRC}^{Prop}} = \int_0^\infty Q_{d,MRC}^{Prob,AWGN} f(\gamma_{MRC}) \, d\gamma_{MRC}, \quad (43)$$

where, $Q_{d,MRC}^{Prob,AWGN}$ can be calculated as follows:

$$\begin{aligned} Q_{d,MRC}^{Prob,AWGN} &\cong Q\left(\frac{\lambda - \mu_{avg}}{\sigma_{avg}}\right) \cdot Q_{d,MRC}^{AWGN}(\lambda/\rho) + \left[1 - Q\left(\frac{\lambda - \mu_{avg}}{\sigma_{avg}}\right)\right] \cdot Q_{d,MRC}^{AWGN}(\rho\lambda) \\ &\cong Q\left(\frac{\lambda - \mu_{avg}}{\sigma_{avg}}\right) \cdot \left[Q_{d,MRC}^{AWGN}(\lambda/\rho) - Q_{d,MRC}^{AWGN}(\rho\lambda)\right] + Q_{d,MRC}^{AWGN}(\rho\lambda) \end{aligned} \quad (44)$$

Given $H_0$, the output of a reliable predictor will be $Q\left(\frac{\lambda - \mu_{avg}}{\sigma_{avg}}\right) \cong 0$, and we get $Q_{fa,MRC}^{Prop} \cong Q_{fa,MRC}(\rho\lambda)$. Hence, $Q_{fa,MRC}^{Prop}$ is lower than the conventional $Q_{fa,MRC}(\lambda)$ in (38). Given $H_1$, the output of a reliable predictor will be $Q\left(\frac{\lambda - \mu_{avg}}{\sigma_{avg}}\right) \cong 1$, and we get $Q_{d,MRC}^{Prob,AWGN} \cong Q_{d,MRC}^{AWGN}(\lambda/\rho)$, so the detection probability $\overline{Q_{d,MRC}^{Prop}}$ is improved compared to the conventional $\overline{Q}_{d,MRC}$ given in (10).

## C. Using SLS

Using SLS, the approximated distribution in (23) is used in (13) and (15) to give [10]:

$$Q_{fa,SLS}(\lambda) \cong 1 - \left[1 - Q\left(\frac{\lambda - N\sigma^2}{\sigma^2\sqrt{2N}}\right)\right]^K, \quad (45)$$

$$Q_{d,SLS}^{AWGN}(\lambda) \cong 1 - \prod_{j=1}^{K}\left[1 - Q\left(\frac{\lambda - N\sigma^2(\gamma_j+1)}{\sigma^2(\gamma_j+1)\sqrt{2N}}\right)\right]. \quad (46)$$

And, $\mu_{avg}$ and $\sigma_{avg}^2$ for SLS can be calculated as:

$$\mu_{avg} = \frac{M}{L}N\sigma^2(\gamma_j+1) + \frac{L-M}{L}N\sigma^2, \quad (47)$$

$$\sigma_{avg}^2 = \frac{M}{L^2}2N\sigma^4(\gamma_j+1)^2 + \frac{L-M}{L^2}2N\sigma^4. \quad (48)$$

Similar to the SLC and MRC cases, (29)-(34), and based on (26) (45) (46), the overall false alarm probability $Q_{fa,SLS}^{Prop}$ and the overall detection probability $\overline{Q_{d,SLS}^{Prop}}$ for the proposed SLS scheme can be expressed as:

$$Q_{fa,SLS}^{Prop} \cong Q\left(\frac{\lambda - \mu_{avg}}{\sigma_{avg}}\right) \cdot [Q_{fa,SLS}(\lambda/\rho) - Q_{fa,SLS}(\rho\lambda)] + Q_{fa,SLS}(\rho\lambda), \quad (49)$$

$$\overline{Q_{d,SLS}^{Prop}} = \int_0^\infty Q_{d,SLS}^{Prob,AWGN} f(\gamma_j)\,d\gamma_j, \quad (50)$$

where, $Q_{d,SLS}^{Prob,AWGN}$ can be calculated as:

$$Q_{d,SLS}^{Prob,AWGN} \cong Q\left(\frac{\lambda - \mu_{avg}}{\sigma_{avg}}\right) \cdot [Q_{d,SLS}^{AWGN}(\lambda/\rho) - Q_{d,SLS}^{AWGN}(\rho\lambda)] + Q_{d,SLS}^{AWGN}(\rho\lambda). \quad (51)$$

Given $H_0$ and using a reliable PU activity predictor, we get $Q\left(\frac{\lambda - \mu_{avg}}{\sigma_{avg}}\right) \cong 0$, then, $Q_{fa,SLS}^{Prop} \cong Q_{fa,SLS}(\rho\lambda)$. Hence, $Q_{fa,SLS}^{Prop}$ is reduced compared to the conventional $Q_{fa,SLS}(\lambda)$ in (45). Given $H_1$, we get $Q\left(\frac{\lambda - \mu_{avg}}{\sigma_{avg}}\right) \cong 1$, then $Q_{d,SLS}^{Prob,AWGN} \cong Q_{d,SLS}^{AWGN}(\lambda/\rho)$, so the detection performance $\overline{Q_{d,SLS}^{Prop}}$ is improved compared to the conventional $\overline{Q}_{d,SLS}$ given in (14).

Therefore, using the same number of CRs, the proposed soft decision CSS has a better performance than the conventional soft decision CSS using SLC, MRC and SLS energy combining schemes.

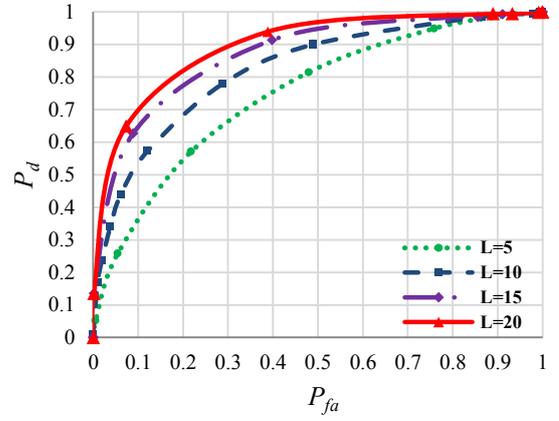

Fig. 2. The ROC curves of the proposed SLC based soft decsion CSS using different $L$ values.

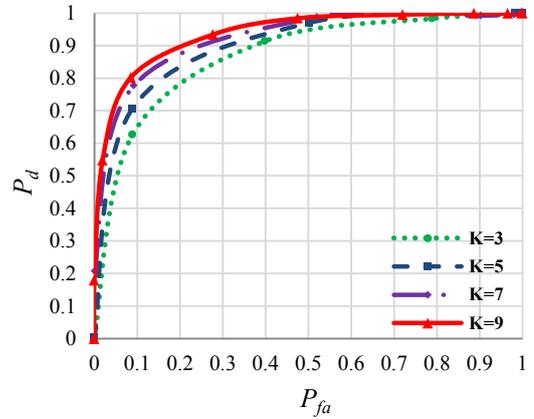

Fig. 3. The ROC curves of the proposed SLC based soft decsison CSS using different number of CRs.

## V. PERFORMANCE EVALUATIONS

As we proved the effectiveness of the proposed scheme over the conventional one via theoretical analysis, in this section we will prove its efficiency via computer simulations. In the conducted simulations, a real communication environment is simulated for the CRN using BPSK and Rayleigh flat fading channel conditions between the PU and all CRs. In addition, noiseless reporting channels are assumed between the FC and the CRs. Monte Carlo computer simulations are used to prove the effectiveness of the proposed scheme using the Receiver Operating Characteristics (ROC) curves. In which, we draw the relationship between the probability of detection ($P_d$) and the Probability of false alarm ($P_{fa}$) for the compared schemes [12].

First, we optimize the performance of the proposed scheme against its critical parameters such as the value of the consecutive energy records $L$ and the number of used CRs using the SLC scheme. Figure 2 shows the ROC curves using different $L$ values, i.e., $L=5, 10, 15$ and $20$ and SNR of -15 dB. The number of used samples $N$ is equal to 1000 samples, and the number of CRs $K$ is equal to 3. From this figure, as the number of stored records $L$ is increased the performance of the proposed scheme is enhanced. Based on the trade-off between performance and complexity, we choose $L=15$ as a sufficient value for the proposed scheme.

In Fig. 3, the number of CRs $K$ is adjusted using SLC, the adjusted value of $L$ ($L=15$), $N=1000$ and SNR=-15 dB. From this figure, $K=7$ is chosen as a sufficient number of CRs.

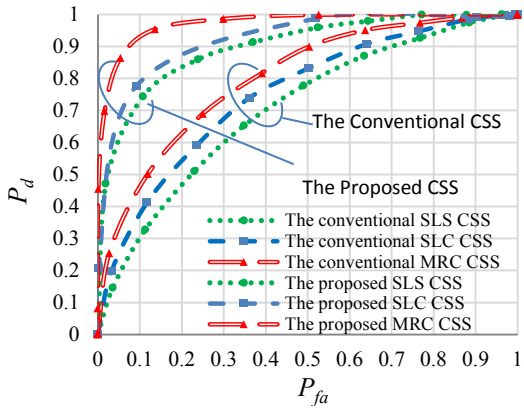

Fig. 4. The performance comparison between the proposed soft decsison CSS and the conventional one.

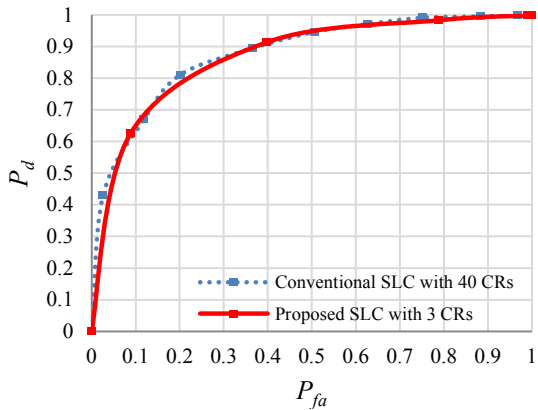

Fig. 5. The complexity comparison between the proposed SLC based soft decsison CSS and the conventional one.

Fig. 4 shows the performance comparisons between the proposed soft decision CSS and the conventional one using different energy combining schemes. In this simulation, we use $L=15$, $K=7$, $N=1000$ and SNR=-15 dB. From this figure, at $P_{fa}$ of 0.1 (which is a rated parameter for CRN design), the proposed algorithm increases the $P_d$ of all combining schemes to be near double the $P_d$ obtained by the conventional soft decision CSS. This is because the proposed scheme takes into account the noise uncertainty effect when making the final decision about the current PU activity, of which function is not provided by the conventional soft decision CSS.

These results confirm the theoretical claims that we proved in the aforementioned section that the proposed scheme has a better performance than the conventional one using the same number of CRs and different energy combining schemes.

## VI. COMPLEXITY ANALYSIS

Fig. 5 shows the complexity comparison between the proposed SLC based soft decision CSS and the conventional SLC based soft decision CSS. In this simulation, we use $L=15$, $N=1000$, and SNR = -15 dB. It is clearly shown that the proposed scheme only using 3 CRs, with a simple averaging process, can achieve the same performance as the conventional one using 40 CRs. This means that the proposed scheme succeeds to reduce the number of used CRs by 92.5 % to obtain the same sensing performance like the conventional method. This can be considered as a significant complexity advantage because using a large number of CRs is not preferred in the CRN design. This is because the sensing time will be increased, which in turn results in reducing the overall CRN throughput. In order to further reduce the complexity of the proposed scheme, we propose that the averaging process is successively performed with time. Therefore, memory and computational complexity requirements are highly relaxed.

## VII. CONCLUSION

In this paper, an enhanced energy detection based soft decision CSS technique has been proposed to enhance the CRN spectrum sensing performance with a low computational complexity. In the proposed scheme, the noise uncertainty effect is taken into account when arriving at a final decision about the current existence of the PU. Current PU status prediction is investigated using a successive PU energy measurements during a specific duration. Based on predicting the current PU status, two dynamic thresholds are toggled to increase the probability of detection and decrease the probability of false alarm. The two thresholds are evaluated using an estimated value of the noise uncertainty factor. Theoretical and simulation analysis proved the high efficiency of the proposed scheme over the conventional soft decision based CSS, in which noise uncertainty effect is not considered. Moreover, the proposed scheme succeeded to reduce the number of used CRs by 92.5 % to obtain the same performance like the conventional one. The proposed CSS is reliably applicable to any communication system including the satellite systems.